\documentclass[%
 reprint,
 amsmath,amssymb,
 aps,
]{revtex4-1}

\usepackage{graphicx}
\usepackage{dcolumn}
\usepackage{bm}

\begin{document}


\title{Pseudo-Jahn-Teller interaction among electronic resonant states of H$_3$}

\author{Patrik Hedvall}
 \email{patrik.hedvall@fysik.su.se}

\author{\AA sa Larson}%
 \email{aasal@fysik.su.se}
\affiliation{%
 Department of Physics, Stockholm University, AlbaNova University Center, SE-106 91 Stockholm, Sweden \\
}%


\date{\today}

\begin{abstract}
We study the electronic resonant states of H$_3$ with energies above the potential energy surface of the H$^+_3$ ground state. These resonant states are important for the dissociative recombination of H$^+_3$ at higher collision energies, and previous studies have indicated that these resonant states exhibit a triple intersection. We introduce a complex generalization of the pseudo-Jahn-Teller model to describe these resonant states. The potential energies and the autoionization widths of the resonant states are computed with electron scattering calculations using the complex Kohn variational method, and the complex model parameters are extracted by a least-square fit to the results. This treatment results in a non-Hermitian pseudo-Jahn-Teller Hamiltonian describing the system. The nonadiabatic coupling and geometric phase are further calculated and used to characterize the enriched topology of the complex adiabatic potential energy surfaces.

\begin{description}
\item[PACS numbers]
May be entered using the \verb+\pacs{#1}+ command.
\end{description}
\end{abstract}

\pacs{Valid PACS appear here}
\maketitle


\section{\label{sec:level1}Introduction}
The Jahn-Teller (JT) effect and the pseudo-Jahn-Teller (PJT) effect are well known examples of vibronic coupling phenomena manifesting conical intersections \cite{jtorg, koppelbook, bers, bers2}. In the JT effect, degenerate electronic states are coupled via nontotally symmetric modes of vibrations. This interaction induces a spitting of the electronic components leaving a symmetry required conical intersection \cite{yark} in the high symmetric nuclear configuration with a singular nonadiabatic coupling and a nontrivial geometric phase \cite{lonhig, berry}. The PJT effect is an extension of the JT effect, where one (or more) nondegenerate electronic state is included, which in turn is coupled to the degenerate JT states. 

As the simplest neutral molecule exhibiting the JT interaction, H$_3$ attracts fundamental interest. The effects of the JT conical intersection in the repulsive ground state of H$_3$ has been studied in the context of reactive scattering of $\textrm{H}+\textrm{H}_2$ \cite{reac}. The set of JT parameters for the series of excited bound Rydberg states have been extracted and analysed using multi-channel-quantum defect theory \cite{jungen, staib}. It has further been shown that dissociative recombination of H$_3^+$ at low collision energies is driven by electron capture into these bound Rydberg states via the vibronic JT interaction \cite{greene}. Hence, the vibronic JT coupling is playing a crucial role in H$_3^+$ dissociative recombination. It can further be mentioned that a similar treatment of low energy dissociative recombination of linear polyatomic ions have shown an analogous importance for the related Renner-Teller effect \cite{hco}.

At higher energies, the electron may be captured by H$_3^+$, forming a doubly excited state of H$_3$, which is energetically embedded in the ionization continuum, \textit{i.e.} an electronic resonant state. Orel \textit{et al.} \cite{orel} used the complex Kohn variational method \cite{complexkohn} to compute four repulsive resonant states of H$_3$ at energies around 4-16 eV above the ionic ground state. Electron capture into these resonant states can explain a high-energy peak observed in the measured dissociative recombination cross-section \cite{mats}. It was also found that three of these electronic resonant states were subject to a triple intersection \cite{orel2}. A corresponding intersection between the bound state of Na$_3$ has been identified as a PJT interaction \cite{meis}.

Both the JT and the PJT interactions are well studied in the case of bound electronic states. In the alkali trimers Li$_3$ \cite{li3}, Na$_3$ \cite{na3}, and K$_3$ \cite{k3}, the low lying $E$ states are well described by a JT model. However, for higher excited $E$ states, a close lying state of $A$ symmetry also couples and the interaction are described by a PJT model. The effects of rotation have also been included in such systems with strong vibronic coupling \cite{rot1}, and the implications of the PJT interaction in Na$_3$ was specifically studied in Ref. \cite{rot2}.

Vibronic interaction among electronic resonant states have been studied before \cite{est, feu1, haxton}, and an explicit treatment of the JT interaction was described in Ref. \cite{feu2}. The basic feature of these models are that the model parameters are allowed to be complex, which in turn renders the interaction potential non-Hermitian and complex symmetric. 

In this study, we have performed electron scattering calculations using the complex Kohn variational method \cite{complexkohn} to investigate the electronic resonant states of H$_3$. We then fit the extracted complex adiabatic potential energy surfaces to a generalized $(E+A)\otimes e$ PJT Hamiltonian for electronic resonant states. We also study the non-adiabatic coupling and geometric phase to characterise the topology of the complex adiabatic potential energy surfaces, generated from the non-Hermitian interaction potential. 

In previous study \cite{roos}, the electronic resonant states of H$_3$ have been estimated using bound state calculations. Here we provide a systematic investigation and application of a generalized PJT model to describe electronic resonant states. The complex PJT model (with associated complex model parameters) present here, allows for direct applications to nuclear dynamics. 

In the following section, the bound state PJT model is summarized and a complex generalisation describing the electronic resonant states is introduced. In section III, the method of extracting the complex model parameters is presented and the electron scattering calculations are described. The results are presented in section IV, where the topology of the complex adiabatic potential energy surfaces are analysed with the non-adiabatic coupling and the geometric phase. This is followed by a brief discussion on possible implications for dynamics and scattering processes. In the appendix, the analytical expression for the complex JT model is presented. Throughout this paper, atomic units are used.

\section{Theoretical model}

We study a non-degenerate electronic state of $A$ symmetry coupled to a doubly degenerate electronic state of $E$ symmetry. These are vibronically coupled via a doubly degenerate vibrational mode of $e$ symmetry, \textit{i.e.} we have a $(E+A)\otimes e$ pseudo Jahn-Teller system. The pair of degenerate vibrational modes that breaks the $D_{3h}$ symmetry are denoted by the coordinates $Q_x$ and $Q_y$, which reduces the symmetry to $C_{2v}$ and $C_{s}$, respectively. These vibrational modes are conveniently represented in polar coordinates as $Q_x=\rho \cos(\phi)$ and $Q_y=\rho \sin(\phi)$, collectively denoted $\textbf{Q}$. Dimensionless mass scaled normal coordinates are used which are related to the displacement coordinates $\Delta r_i$, \textit{i.e.} the bond stretching coordinate relative the equilibrium configuration of  $r=1.65$ a$_0$ of the H$_3^+$ ion, where $Q_x=\frac{f}{\sqrt{3}}(2\Delta r_1-\Delta r_2 - \Delta r_3)$, and $Q_y=f(\Delta r_2-\Delta r_3)$ and $f=2.639255$ a$_0^{-1}$ is a constant \cite{mistrik00}. The additional totally symmetric vibrational mode, $Q_s$ is kept frozen at $Q_s=0$. We are here interested in the conical intersection induced by the symmetry breaking and since the $Q_s$ mode preserves the $D_{3h}$ symmetry, it is excluded in this study. The derived Hamiltonian can, however, be generalized to also include the symmetric mode.

\subsection{PJT for electronic bound states}
The $3 \times 3$ matrix Hamiltonian describing the electronically bound PJT system can be expressed, in a diabatic representation, as $\textbf{H}=\textbf{I}\widehat{T}_N +\textbf{V}^d$, where $\widehat{T}_N$ is the nuclear kinetic energy operator, $\textbf{I}$ the identity matrix and $\textbf{V}^d$ is the diabatic potential energy matrix, with elements 

\begin{equation}
V_{nm}^d=\langle \psi_n \vert H_{el}\vert \psi_m\rangle \quad \quad n,m=E_x,E_y, A. \label{v1}
\end{equation}
Here $\vert \psi_{E_x}\rangle$, $\vert \psi_{E_y}\rangle$ denotes the real valued, doubly degenerate electronic components transforming as the vibrational coordinates and $\vert \psi_A\rangle$ denotes the totally symmetric real valued electronic state. $H_{el}$ is the electronic Hamiltonian of the system.

The PJT diabatic Hamiltonian can be derived by a Taylor expansion of \eqref{v1} in the vibrational coordinates around the symmetric $D_{3h}$ configuration and the non-zero expansion coefficients are determined by symmetry considerations \cite{veil}. In second order JT and first order PJT, the elements of the $3\times 3$ diabatic potential governing the nuclear motion in real representation are

\begin{equation}
V_{nm}^d =\Big(\varepsilon_n + \frac{1}{2}\omega \rho^2\Big)\delta_{nm} + k J_{nm}^k + g J_{nm}^g+ \alpha J_{nm}^{\alpha} \label{jt1}.
\end{equation}
The first term is diagonal and corresponds to a harmonic approximation of the uncoupled states, where $\boldsymbol{\varepsilon}=\big(\varepsilon_A,\varepsilon_E,\varepsilon_E \big)$ denotes the energy of the electronic states in the $D_{3h}$ configuration. The coupling constants $k$ and $g$ together with the coupling matrices

\begin{equation}
\textbf{J}^k=\rho \begin{pmatrix} 0 & 0 & 0 \\
0 &  \cos (\phi)  &   \sin (\phi)  \\
0 &   \sin (\phi) & -\cos (\phi)  
\end{pmatrix} \label{J1}
\end{equation}
and

\begin{equation}
\textbf{J}^g =\rho^2  \begin{pmatrix} 0 & 0 & 0 \\
0 &    \cos(2\phi) &  - \sin(2\phi) \\
0 &  - \sin(2\phi) & -  \cos(2\phi) 
\end{pmatrix} \label{J2}
\end{equation}
describe the JT interaction among the degenerate $E$ components, in first and second order, respectively. The PJT interaction with the $A$ state is described by the coupling constant $\alpha$ and the coupling matrix

\begin{align}
\textbf{J}^{\alpha}&=\rho \begin{pmatrix} 0 &   \cos (\phi) & - \sin (\phi) \\
  \cos (\phi) & 0 & 0 \\
- \sin (\phi) & 0 & 0 
\end{pmatrix} \label{J3}.
\end{align}
 Often the bound state diabatic PJT (and JT) potential is expressed in a complex (Hermitian) representation to reveal the symmetry of the system \cite{veil}. We will later introduce a complex (non-Hermitian) generalisation of the PJT potential describing electronic resonant states. It is therefore convenient at this stage to express the bound state potential in a real representation.

Adiabatic potential energy surfaces are obtained by diagonalising \eqref{jt1}, and will be denoted $V_n(\textbf{Q})$ in rising energy order. In the direction of the $C_{2v}$  preserving coordinate $Q_x$ (corresponding to $\phi=0,\pi$ or $Q_y=0$), the adiabatic potential energy surfaces attain a simple form, 
\begin{align}
V_{1/3}(Q_x) &= \frac{1}{2}(\varepsilon_E+\varepsilon_A) + \frac{1}{2}\omega Q_x^2 + \frac{1}{2}(kQ_x + g Q_x^2) \label{vpm}\\
&\pm \sqrt{\frac{1}{4}\big(\varepsilon_A - \varepsilon_E- k Q_x-gQ_x^2\big)^2 + (\alpha Q_x)^2} \nonumber \\
V_{2}(Q_x) &= \varepsilon_E + \frac{1}{2}\omega Q_x^2 - k Q_x - g Q_x^2 \label{v0} \end{align}
convenient for fitting to \textit{ab initio} treatments. Expansion of $V_{1/3}$ shows that in linear order only $k$ is present, while $\omega$, $g$ and $\alpha$ comes in second order in $Q_x$. A second order coupling to the \textit{A} state can be included in the diabatic potential, but it appears in third order in the adiabatic potentials. Therefore the expansion \eqref{jt1} is referred to as the second order PJT model. 

For a non-zero linear coupling $k$, a conical intersection is present among the degenerate $E$ states when $V_1=V_2$ at $\rho=0$. The geometric phase associated with the conical intersection is $\pi$, which for the nuclear dynamics give rise to half-odd integer rotational quantum numbers \cite{lonhig}. Three additional conical intersections can be found when $V_1=V_2$ at a critical radius $\rho_c$. For a path encircling all four intersections, i.e. for a fixed $\rho>\rho_c$, the total geometric phase sum to an even multiple of $\pi$, with a cancellation of the geometric phase effects. Such suppression of the geometric phase has been found and analysed in systems like Na$_3$ \cite{meis, meis2} with strong PJT coupling relative linear JT coupling.

\subsection{PJT for electronic resonant states}
The electronic resonant states are modelled as discrete bound states interacting with a continuum of scattering states \cite{est}. In the Feshbach projection operator formalism \cite{fesh}, two complementary projection operators $Q$ and $P$ are introduced, which partition the Hamiltonian $H$ and the wavefunction $\Psi=Q\Psi + P\Psi$ into the discrete states and continuum states. The Q-space portion we associate with the PJT system introduced above, and define the operator

\begin{equation}
Q=\vert \psi_{E_x} \rangle\langle \psi_{E_x} \vert + \vert \psi_{E_y} \rangle \langle \psi_{E_y} \vert+ \vert \psi_A \rangle \langle \psi_A \vert
\end{equation}
which projects onto the three discrete diabatic PJT electronic states. Its complementary operator projects onto a set of continuum states, 

\begin{equation}
P=\int d\boldsymbol{\epsilon} \ \vert \psi_{\boldsymbol{\epsilon}} \rangle \langle \psi_{\boldsymbol{\epsilon}} \vert
\end{equation}
where $\boldsymbol{\epsilon}=\epsilon \widehat{\omega}$ denotes the energy $\epsilon$ and direction $\widehat{\omega}$ of the ejected electron. Imposing purely outgoing boundary conditions on $P\Psi$, an effective Hamiltonian governing the Q-space portion can be expressed as \cite{est}
\begin{equation}
\mathcal{H}_{eff}=QHQ+QHP G_P^+PHQ. \label{heff}
\end{equation}
Here $G_P^+$ is the outgoing Green's function analytically continued to the complex energy plane,

\begin{equation}
G_P^+=\frac{1}{E-PHP+i\eta}.
\end{equation}
$\mathcal{H}_{eff}$ is an effective, energy dependent Hamiltonian describing the Q-space portion of the system.

The local complex model (or the Boomerang model) \cite{boomerang, hazi1} is a well established treatment of the effective Hamiltonian $\mathcal{H}_{eff}$ \eqref{heff}, where a number of simplifying assumptions are made, such that the potential term can be expressed sorely in terms of the nuclear vibrational modes $\textbf{Q}$. In this approximation, the energies of the resonant states are assumed to be high enough such that the open vibrational states of the target forms a complete set \cite{bieniek}. The coupling elements between the discrete states and the continuum are assumed to be independent of the ejected electron energy. Expressed in the basis of the diabatic electronic states, the effective diabatic Hamiltonian in the local complex model can be written \cite{est}

\begin{equation}
\mathcal{H}_{nm}^d = \widehat{T}_N \delta_{nm} + V_{nm}^d(\textbf{Q})+\Delta_{nm}(\textbf{Q}) -\frac{i}{2} \Gamma_{nm}(\textbf{Q}), \label{hjt}
\end{equation}
where $V_{nm}^d(\textbf{Q})$ is the bound state potential matrix given by \eqref{jt1}, which includes the direct vibronic interactions. The terms $\Delta_{nm}(\textbf{Q})$ and $\Gamma_{nm}(\textbf{Q})$ in \eqref{hjt} manifests the continuum interaction and includes both the decay (autoionization) mechanism and an indirect coupling mechanism \cite{hazi83}, allowing the electron to hop from one discrete state to the other via the continuum. $\Delta_{nm}(\textbf{Q})$ is called the potential energy shift and is often incorporated into $V_{nm}^d(\textbf{Q})$ and $\Gamma_{nm}(\textbf{Q})$ is referred to as the diabatic width. If the three resonant state were uncoupled the real part and the imaginary part of \eqref{hjt} would correspond to the resonance parameters (energy and width) entering the Breit-Wigner formula, \textit{i.e.} equations \eqref{bw1} and \eqref{comxpot} below. In this case of coupled resonant states the Breit-Wigner parameters are associated with the complex eigenvalues of \eqref{hjt}, \textit{i.e.} the complex adiabatic potential energy surfaces.

Since the potential energy shift, $\Delta_{nm}(\textbf{Q})$ can be expressed in terms of $\Gamma_{nm}(\textbf{Q})$ \cite{est}, it possesses the same symmetry. The diabatic width, $\Gamma_{nm}(\textbf{Q})$ can in turn be expressed as
 
\begin{equation}
\Gamma_{nm}(\textbf{Q})= 2\pi  \int d \widehat{\omega} \ \langle \psi_n \vert H_{el}\vert \psi_{\boldsymbol{\epsilon}}\rangle \  \langle \psi_{\boldsymbol{\epsilon}} \vert H_{el}\vert \psi_m \rangle,  \label{gam}
\end{equation}
where the coupling elements between the discrete states and the continuum are evaluated at the coordinate dependent resonance energy \cite{est} and integrated over the solid angle of the ejected electron with energy $\epsilon$.

Both the electronic Hamiltonian $H_{el}$ and the projection operator onto the continuum states $\int d \omega \ \vert \psi_{\boldsymbol{\epsilon}} \rangle \langle \psi_{\boldsymbol{\epsilon}} \vert$ in \eqref{gam} are invariant under symmetry transformations in $D_{3h}$, \textit{i.e} they belong to the $A$ irreducible representation. This implies that the matrix elements $\Gamma_{nm}(\textbf{Q})$ and $\Delta_{nm}(\textbf{Q})$ follow the same symmetry transformations as the bound state potential elements $V_{nm}^d(\textbf{Q})$. An expansion of $\Gamma_{nm}(\textbf{Q})$ and $\Delta_{nm}(\textbf{Q})$ around $\rho=0$ will therefore attain the same functional form as $V_{nm}(\textbf{Q})$, but with other values of the expansion parameters. Thus, we allow for the set of PJT parameters $\lbrace \varepsilon_E, \varepsilon_A, \omega, k, g, \alpha \rbrace$ to be complex, where the real parts correspond to the expansion coefficients for the energy potential $V_{nm}(\textbf{Q})+\Delta_{nm}(\textbf{Q})$, while the imaginary parts act as expansion coefficients for the width $\Gamma_{nm}(\textbf{Q})$, \textit{i.e.} the expansion of $\Gamma_{nm}(\textbf{Q})$ around the $D_{3h}$ configuration can be expressed as

\begin{align}
-\frac{1}{2}\Gamma_{nm}(\textbf{Q})&=\Big(\textrm{Im}(\varepsilon_n) + \frac{1}{2}\textrm{Im}(\omega) \rho\Big)\delta_{nm} \\
&+ \textrm{Im}(k) J_{nm}^k + \textrm{Im}(g) J_{nm}^g+ \textrm{Im}(\alpha) J_{nm}^{\alpha}\nonumber \label{jt2}
\end{align}
with the coupling matrices given by \eqref{J1}, \eqref{J2} and \eqref{J3}.

The PJT Hamiltonian \eqref{hjt} governing the electronic resonant states is now complex, symmetric and non-Hermitian \cite{feu2}. In the vicinity of $\rho=0$ it is fully characterised by the complex parameters $\lbrace \varepsilon_E, \varepsilon_A, \omega, k, g, \alpha \rbrace$, which implies that the expressions for the potential energy surfaces \eqref{vpm} and \eqref{v0} still apply, but now with complex parameters. This is in complete analogy with the treatment of the JT interaction among electronic resonant states developed by Feuerbacher \textit{et al.} \cite{feu2, feu1}.

\section{Extracting model parameters for H$_3$}
In this section we describe how the complex model parameters entering the diabatic potential are extracted by fitting the complex adiabatic potential energy surfaces to results from fixed nuclei electron scattering calculations on the H$_3^++e$ system.

\subsection{Electron scattering calculations}
The H$_3^+$ molecular system has $D_{3h}$ symmetry and dominant configuration $1a_1^2$. The lowest bound states of H$_3$ responsible for the JT-effect have the configuration $1a_1^2 1e^1$, where the $1e$-molecular orbital corresponds to the $2p \pi$ orbital in the united atom limit \cite{kokou}. The lowest electronic resonant states of H$_3$ have doubly excited configurations $1a_1^1 1e^2$ \cite{orel2}. When the molecule distorts to $C_{2v}$ symmetry, the components of the $1e$ orbital split into $2a_1$ and $1b_2$.  Different electronic resonant states are formed with dominant configurations corresponding to $1a_1^1 2a_1^2$,   $1a_1^1 1b_2^2$ or $1a_1^1 2a_1^1 1b_2^1$. The spins of the two electrons in two different components of the $1e$ orbital may be singlet- or triplet coupled. As described by Orel { \it et al.}   \cite{orel2}, there is a conical intersection between the lower pairs of $^2A_1$ and $^2B_2$ states as well as a strong avoided crossing between the two states of $^2A_1$ symmetry. These are the electronic resonant states that are studied here. 

In order to describe the electronic resonant states, we have performed electron scattering calculations on the H$_3^+$ + e system using the complex Kohn variational method \cite{complexkohn}. These calculations explicitly includes the scattering wave functions in the trial wavefunction as well as terms containing square-integrable configuration state functions of the H$_3$ system. The parameters of the trial wave function are optimized for a given scattering energy of the electron using the complex Kohn functional. Then, from the scattering matrix, the energy positions and autoionization widths of the resonant states can be determined at fixed nuclear geometries.

The target H$_3^+$ wave function is described using a hydrogen basis set of $(10s,5p)$ contracted to $[7s,5p]$. Natural orbitals of the target are determined using a full configuration interaction calculation. This is followed by an augmentation with $(3s,3p)$ at the center of mass. Using the natural orbitals, the target wave function is constructed from a multi-reference configuration interaction calculation, where excitations of the two electrons among six orbitals are included. The electron scattering calculations are carried out by including partial waves with angular momentum $l \leq 4$ and $|m| \leq 4$.

The eigenphase sum $\delta(E)$, which is directly related to the scattering matrix, is convenient to analyse the scattering resonances \cite{hazi79}. Both the energy and width of the resonant state can be extracted by fitting the eigenphase sum to the Breit-Wigner formula. However, for overlapping resonances like the ones in H$_3$, extracting parameters using the time-delay is preferable \cite{timedelay1, timedelay2}. The time-delay can be obtained from the derivative of the eigenphase sum

\begin{equation}
\frac{d\delta}{dE}=\sum_n \frac{\gamma_n/2}{(E-\epsilon_n)^2+(\gamma_n/2)^2} + \frac{d\delta_{bg}}{dE}, \label{bw1}
\end{equation}
where the resonance parameters $\epsilon_n$ and $\gamma_n$ denote the energy position and the width of resonance $n$ for a fixed nuclear geometry $\textbf{Q}$. The resonance paramaters are extracted by fitting the time-delay to the formula above, where the derivative of the background phase shift is taken as a constant. An adiabatic potential describing the resonant state $n$ can in turn be expressed in terms of the resonance parameters in the following complex form \cite{boomerang, kapur}

\begin{equation} V_n(\textbf{Q})=V_{ion}(\textbf{Q})+\epsilon_n(\textbf{Q})-\frac{i}{2}\gamma_n(\textbf{Q}) \label{comxpot},
\end{equation} 	
where $V_{ion}$ is the energy of the target H$_3^+$ ion. The results from the electron scattering calculations, \textit{i.e.} the real- and imaginary parts of $V_n$ for a number of nuclear geometries are displayed with dots in Figures 1, 2 and 3 below.

\subsection{PJT parameters}
The complex model parameters are extracted by fitting the adiabatic potentials given by equations \eqref{vpm} and \eqref{v0} to the resonance parameters obtained from the electron scattering calculations given by equation \eqref{comxpot} along the C$_{2v}$ preserving coordinate $Q_x$ (\textit{i.e.} keeping $Q_y=0$). In C$_{2v}$ symmetry the degenerate $E$ component splits into components of $A_1$ and $B_2$ symmetries. By performing the electron scattering calculations separately in $A_1$ and $B_2$ symmetries we circumvent the difficulty of fitting overlapping resonances lying very close in energy, \textit{i.e.} in the vicinity of the conical intersection.

The model parameters are extracted by the least square fit method over the range $Q_x=\pm 0.5$ and the results are shown in Fig. 1, where the dots are the results from the electron scattering calculations and the lines are the fitted potentials. The extracted complex PJT parameters are presented in the first column of Table 1.

In Fig. 2, the degenerate components subject to the conical intersections are shown in more detail and the PJT model (solid line) is compared to a fit of a pure JT model (dashed line) (see Appendix). The fitted parameters from the pure JT model are presented in the third column of Table 1. These are obtained by setting $\alpha=0$ in the potentials \eqref{vpm} and \eqref{v0} and excluding the $A$ state from the fit. Further, in C$_{2v}$ symmetry, the JT potentials can be written as $V_{1/2}(Q_x)= \varepsilon_E + \frac{\omega}{2}Q_x^2  \pm (k Q_x + g Q_x^2)$ such that the real and imaginary parts are separable and fits can be done separately for $\textrm{Re}(V_{1/2})$ and $\textrm{Im}(V_{1/2})$  \cite{feu2}. In contrast, for the PJT model both the real and the imaginary parts of the parameters need to be fitted simultaneously.

\begin{figure}[!ht]
\includegraphics[scale=0.41]{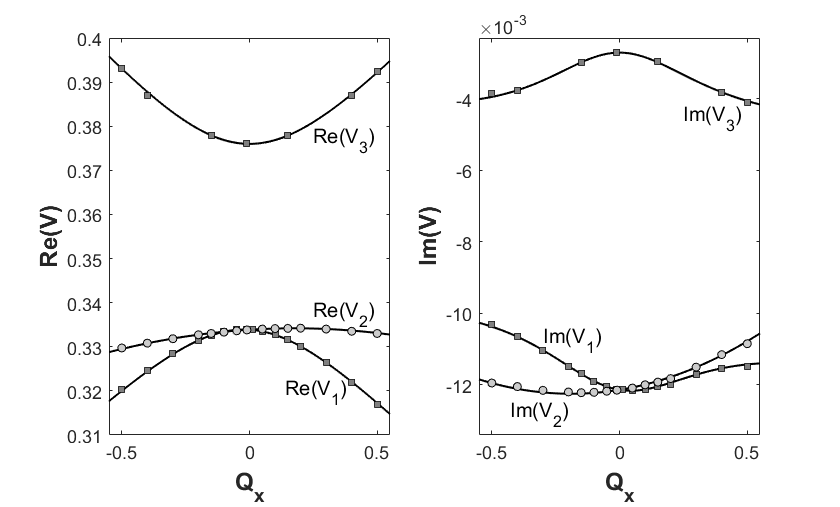}
\centering
\caption{The results from the electron scattering calculations (dots) and the fitted complex adiabatic potentials (solid lines) along the $Q_x$ coordinate when $Q_y=0$. The coordinates are presented in dimensionless units and the potential energy in Hartree. The left panel shows the real part of the potentials associated with the energies and the right panel shows the imaginary part associated with the autoionization widths. Along the C$_{2v}$ preserving coordinate $Q_x$ the state $V_2$ transform as $B_2$ while the states $V_{1/3}$ transform as $A_1$. }
\end{figure}

\begin{figure}[!ht]
\includegraphics[scale=0.39]{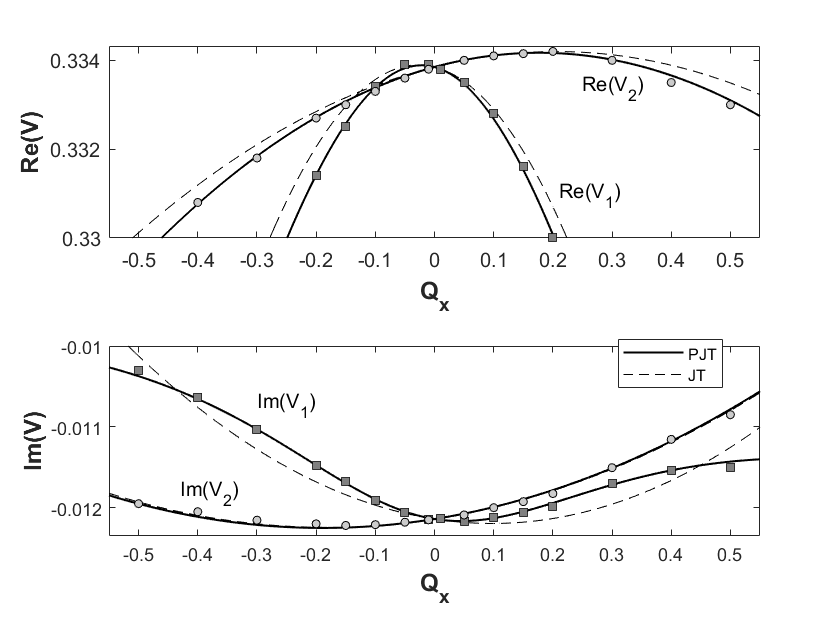}
\centering
\caption{A detailed figure of the lower potential energy surfaces $V_1(Q_x, Q_y=0)$ and $V_2(Q_x, Q_y=0)$, for the complex PJT model (solid lines) and the complex JT model (dashed lines). The dots are the results from the electron scattering calculations. The coordinates are presented in dimensionless units and the potential energy in Hartree. Notice that the real parts of the potential energy surfaces crosses at two points.}
\end{figure}

\begin{table*}[t]
	\centering
	\setlength\extrarowheight{3pt}
	\begin{tabular}{cccc} 
		\toprule
		& PJT model (second order) $\quad$ & $\quad \quad$ PJT model (third order) $\quad$  & $\quad$  JT model (second order) $\quad \quad$       \\
		\\[-1em]
		\hline 
		$\quad \ \varepsilon_E \    \quad \quad$  &  $\ \ 0.3339-i0.0121 \quad$ & $\quad \ \ 0.3339-i0.0121$ & $\ \ \quad 0.3339-i0.0121 \quad  $  \\ 
		$\quad \ \varepsilon_A \  \quad \quad$ &  $ \ \ 0.3760-i0.0027 \quad $& $\quad \ \ 0.3760-i0.0027$ & $ \quad \ \ --- \quad $  \\ 
		$\quad \omega \quad \quad $ & $-0.0031+i0.0019 \quad $ & $\quad -0.0033 + i0.0020$ & $\quad -0.0741+i0.0089 \quad $               \\
		$\quad k  \quad \quad $ &  $-0.0037 - i 0.0012 \quad$  & $\quad-0.0036 - i 0.0011$ & $\quad -0.0034 - i 0.0011 \quad $  \\ 
		$\quad g  \quad \quad $ & $\ \ 0.0085 - i 0.0021 \quad$ & $\quad \ \ 0.0086 - i 0.0021$ & $\quad \ \ 0.0268 + i 0.0014 \quad $       \\
		$\quad \alpha  \quad \quad $ &  $ \ \ 0.0627 + i 0.0018 \quad$ &$\quad \ \ 0.0627 + i 0.0018$ &  $ \quad --- \quad  $  \\
$\quad \beta \quad \quad  $ &  $ \ \ --- \quad$ & $\quad \ \ 0.0011 + i 0.0000$ &  $\quad --- \quad $  \\    
$\quad \nu  \quad \quad $ &  $ \ \ --- \quad$ &$\quad-0.0005 - i 0.0003$ &  $\quad --- \quad $  \\    
$\quad \mu  \quad \quad $ &  $ \ \ --- \quad$ &$\quad -0.0006 - i 0.0004$ &  $ \quad --- \quad $  \\		
\\ [-1em]
		\toprule
	\end{tabular}
	\caption{The fitted parameters for the complex PJT model in second order and third order are presented in the first and second columns, respectivly. The fitted parameters for the complex JT model in second order are presented in the third column. Since dimensionless coordinates are used in this study, the fitted parameters have units of Hartree.}
\end{table*}

The slices of the potential energy surfaces are well described by the complex PJT model, with a strong PJT coupling $\alpha$ and small linear JT coupling $k$. In addition to the crossing at $\rho=0$ where $V_1=V_2$ and both the real and imaginary parts of the complex adiabatic potentials intersects, the real part of the energy surfaces also crosses at $Q_x \approx -0.1$, as indicated by the electron scattering calculations. This is captured by both the complex PJT and the complex JT models.

For comparison, we also fit the results to a higher order PJT model  \cite{veil} and these parameters are presented in the second column of Table 1. This fit includes a second order coupling to the symmetric $A$ state described by the parameter $\beta$, and a third order coupling between the $E$ states described by the parameter $\mu$. Also, a third order term $\nu Q_x^3$ is included in the diagonal term in \eqref{jt1}. This model is refered to as third order PJT since the parameters $\lbrace \beta, \nu, \mu\rbrace$ comes in third order in the adiabatic potentials. The inclusion of these terms only gives a minor shift for the parameters obtained from the second order PJT model fit, which indicates that the second order complex PJT model is sufficient within this range of $Q_x$. It should be mentioned that the uncertanties in the parameters also stems from the fitting of the time-delay and the uncertanties in the \textit{ab initio} electron scattering calculations \cite{uncert}. We have not done a systematic investigation of these uncertanties.

\begin{figure}[!ht]
\includegraphics[scale=0.39]{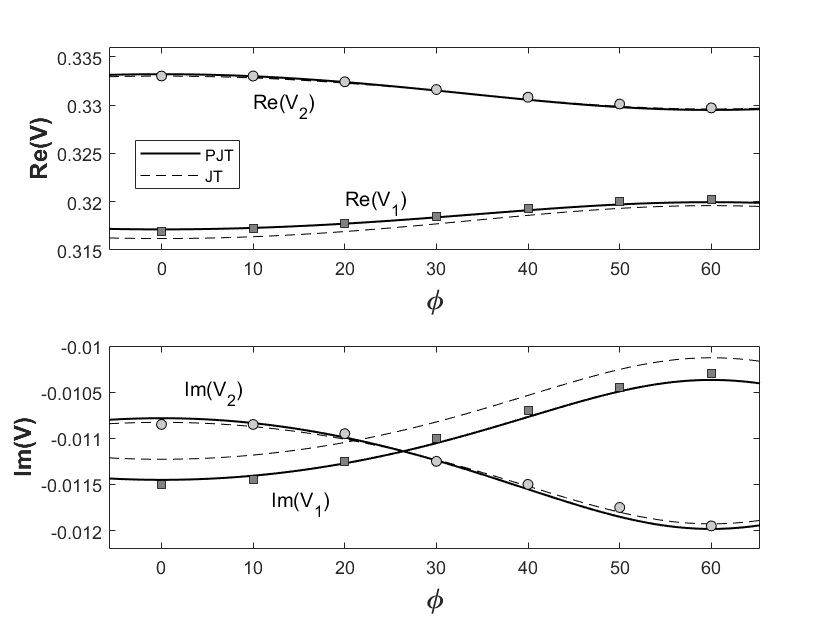}
\centering
\caption{The results (dots) from the electron scattering calculations in $C_s$ symmetry compared to the model behaviour for the extracted parameters from the $C_{2v}$ fits. Here, $\rho=0.5$ is fixed while the angle $\phi$ is varied. The solid lines show the results from the PJT model, while the dashed lines are the ones from the JT model. The angle $\phi$ is presented in degrees and the potential energy in Hartree.}
\end{figure}

In order to investigate the behavior of the models for $Q_y \neq 0$, we have performed electron scattering calculation in C$_s$ symmetry. The parameters obtained from the fit of the potentials along the C$_{2v}$ slice,  are inserted in the expressions for the complex adiabatic potential energy surfaces and compared to the results from the C$_s$ calculations. A comparison for a fixed $\rho=0.5$ and varied angles $\phi$ is shown in Fig. 3, where it shows that the angle dependence is well captured by the PJT model. The scattering calculation for the higher lying $A$ state show only a minor variation in $\phi$, which is captured by the PJT model, but is not displayed in the figure. However, in C$_s$ symmetry we cannot extract accurate resonance parameters for nuclear coordinates closer to the conical intersection, since the resonant states that now have the same symmetry are highly overlapping.

\section{Topology of the resonant states}
 In this section we study the topology associated with the two lowest complex adiabatic potential energy surfaces in the complex PJT model, \textit{i.e.} the interaction among the degenerate $E$ components. To complement the analysis, some properties of the complex JT model, where analytic expressions are available, are given in the appendix. Even though the JT model gives a poor quantitative description of the states, it captures (qualitatively) the characteristic features of the topology. For the complex PJT model we resort to numerical evaluation. 

The transformation between diabatic and adiabatic representation is achieved with the eigenvector matrix $\textbf{T}$, which diagonalises $V_{nm}^d+\Delta_{nm}-i\Gamma_{nm}/2$. Even though such transformation does not exist in a strict sense for polyatomic molecules \cite{meadtru}, it is a well studied approximation. As a consequence of the non-hermiticity of the potential, the dual eigenvector matrix $\tilde{\textbf{T}}$ must also be considered, corresponding to biorthogonal eigenvectors \cite{brody}. Since we originally choose a real representation the dual eigenvector matrix is simply defined $\tilde{\textbf{T}}=\textbf{T}^{T}$, as transpose only and not hermitian conjugate. Transforming the diabatic Hamiltonian \eqref{hjt} to an adiabatic representation, we obtain

\begin{equation}
\mathcal{H}^a = \textbf{T}^{T} \mathcal{H}^d \textbf{T} = \textbf{I}\widehat{T}_N+\textbf{V}^{a} +  \widehat{\boldsymbol{\Lambda}}, 
\end{equation}
where $\textbf{V}^a$ is a diagonal matrix with the complex adiabatic potential energy surfaces $V_n(\textbf{Q})$ as elements, which reduces to equations \eqref{vpm} and \eqref{v0} with complex parameters for $Q_y=0$. 

The complex adiabatic potential energy surfaces $V_{n}(\textbf{Q})$, are presented in Figures 4 and 5. The black dots indicates point of intersections where both the real and imaginary parts of the lower two adiabatic surfaces intersect. In addition to the central intersection at $\rho=0$, there are six outer intersections at $\rho_c=0.107$, laying pairwise symmetric at angles $\phi_1=60\pm 8.68^o$, $\phi_2=180\pm 8.68^o$ and $\phi_3=300\pm 8.68^o$. The solid black line shows seams where either $\textrm{Re}(V_1)=\textrm{Re}(V_2)$ or $\textrm{Im}(V_1)=\textrm{Im}(V_2)$.

\begin{figure}[!ht]
\includegraphics[scale=0.4]{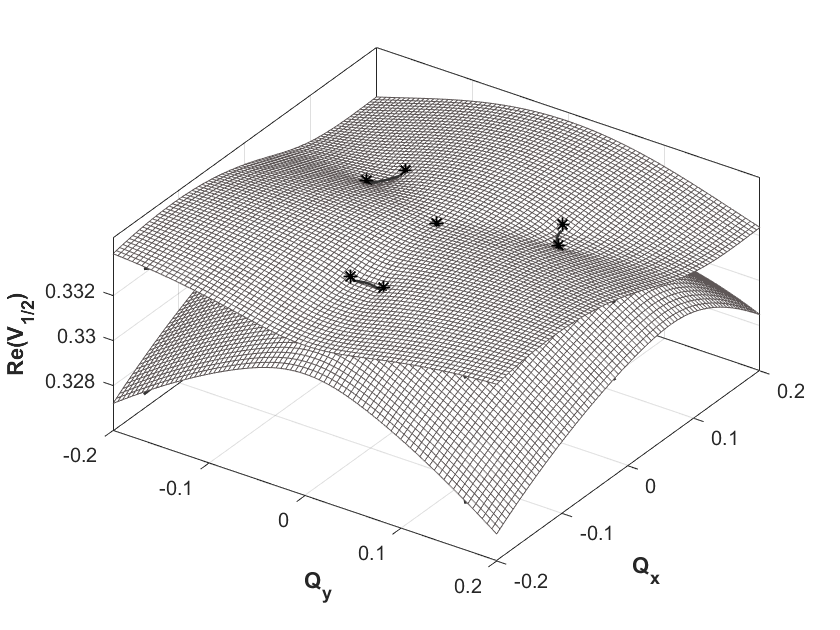}
\centering
\caption{The real part of the two lower complex adiabatic potential energy surfaces $\textrm{Re}(V_1)$ and $\textrm{Re}(V_2)$. The complex conical intersection and the exceptional points are marked by dots and the solid line marks the seams where $\textrm{Re}(V_1)=\textrm{Re}(V_2)$. The coordinates are presented in dimensionless units and the potential energy in Hartree.}
\end{figure}

\begin{figure}[!ht]
\includegraphics[scale=0.4]{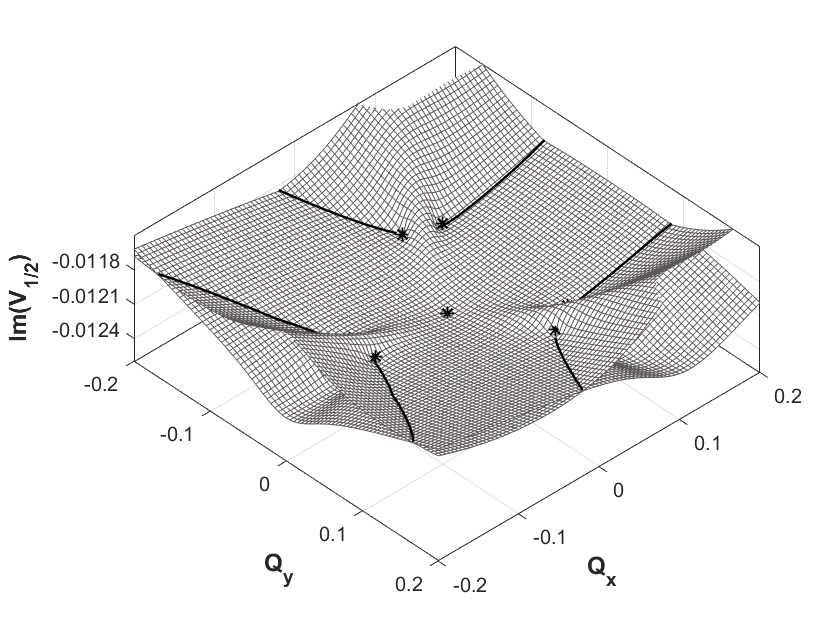}
\centering
\caption{The imaginary part of the two lower complex adiabatic potential energy surfaces $\textrm{Im}(V_1)$ and $\textrm{Im}(V_2)$, which are related to the autoionization widths. The complex conical intersection and the exceptional points are marked by dots and the solid line marks the seams where $\textrm{Im}(V_1)=\textrm{Im}(V_2)$. The coordinates are presented in dimensionless units and the potential energy in Hartree. }
\end{figure}

The complex non-adiabatic coupling operator $\widehat{\Lambda}_{nm}$ carries both the conventional bound state non-adiabatic coupling and the second order interaction through the continuum. It can be expressed as \cite{koppelbook}

\begin{equation}
\Lambda_{nm}= \sum_k \textbf{F}_{nk} \textbf{F}_{km} + \nabla_{\textbf{Q}} \cdot \textbf{F}_{nm} +2 \textbf{F}_{nm} \cdot \nabla_{\textbf{Q}},
\end{equation}
completely characterised by the first derivative non-adiabatic coupling elements

\begin{equation}
\textbf{F}_{nm} = \sum_{k} T^{T}_{nk} \nabla_{\textbf{Q}} T_{km}. \label{F1} \\
\end{equation}
which is a matrix with vector valued elements. The non-adiabatic coupling is one of the more important quantities in the theory of non-adiabatic reactions. It measures the validity of the Born-Oppenheimer approximation, meeting its extreme at conical intersections where it may diverge, marking a total break down of the Born-Oppenheimer approximation. The circulation of the non-adiabatic coupling in the space of the nuclear coordinates provides an identification of the intersection \cite{yark}, \textit{i.e.} via the geometric phase (Berry phase) \cite{lonhig, berry}. For a closed curve $C$ in the coordinate space the geometric phase can be evaluated
 
\begin{equation}
\tau = i \oint_C  \textbf{F}_{nn}^s \ d\textbf{Q},
\end{equation}
where $\textbf{F}_{nn}^s$ refers to the single valued version of \eqref{F1} (see appendix).

Since analytical expressions of the eigenstates are not available for the PJT model, we evaluate the non-adiabatic coupling and the geometric phase numerically \cite{wagner}. This involves a gauge smoothing to generate continuous and single valued biorthogonal eigenstates over the coordinate space. In Figures 6 and 7, the real and imaginary parts of the off-diagonal elements of the first derivative non-adiabatic coupling in $\phi$ direction are shown. At $\rho=0$ the real part $\textrm{Re}(F_{12})$ diverges while the imaginary part does not. The geometric phase associated with the central intersection evaluates to $\tau=\pi$, similarly as in the bound state situation with a JT or PJT conical intersection. Thus, we can conclude that the central intersection is indeed a conical intersection with the conventional geometric phase of $\pi$.

Encircling one of the outer intersections, the states are interchanged and the geometric phase evaluates to $\pi/2$. Thus, two loops around the intersection are needed to attain the conventional sign change. This is a consequence of the non-hermiticity of the system, and these outer intersections are further identified as exceptional points \cite{heiss}. Encircling all 7 intersections, the total geometric phase evaluates to $\tau=2\pi$ and similarly to the bound state situation the geometric phase effect is cancelled out.

\begin{figure}[!ht]
\includegraphics[scale=0.37]{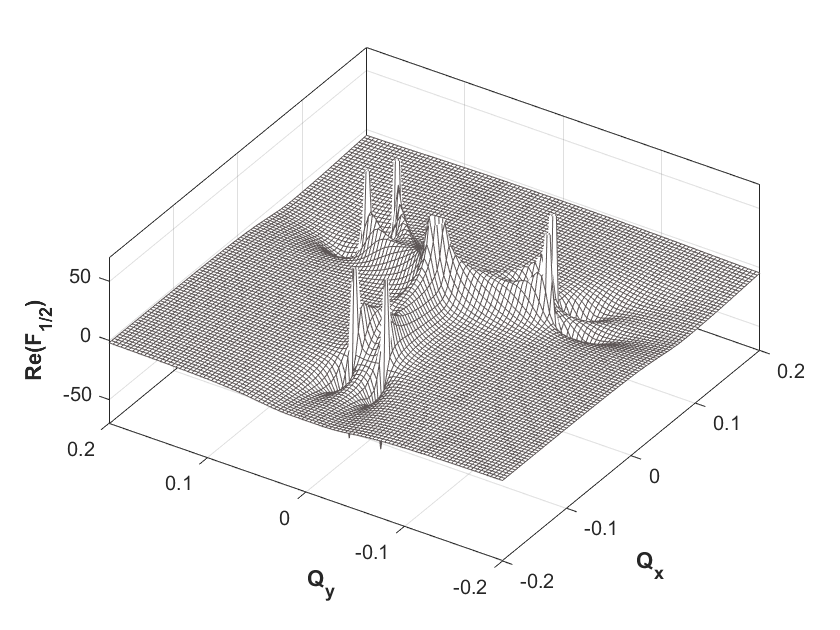}
\centering
\caption{The real part of the first derivative non-adiabatic coupling in $\phi$ direction between the two lower electronic resonant states. At the geometries of the complex conical intersection and the exceptional points the real part diverges. The coordinates and the non-adiabatic coupling are presented in dimensionless units.}
\end{figure}
\begin{figure}[!ht]
\includegraphics[scale=0.37]{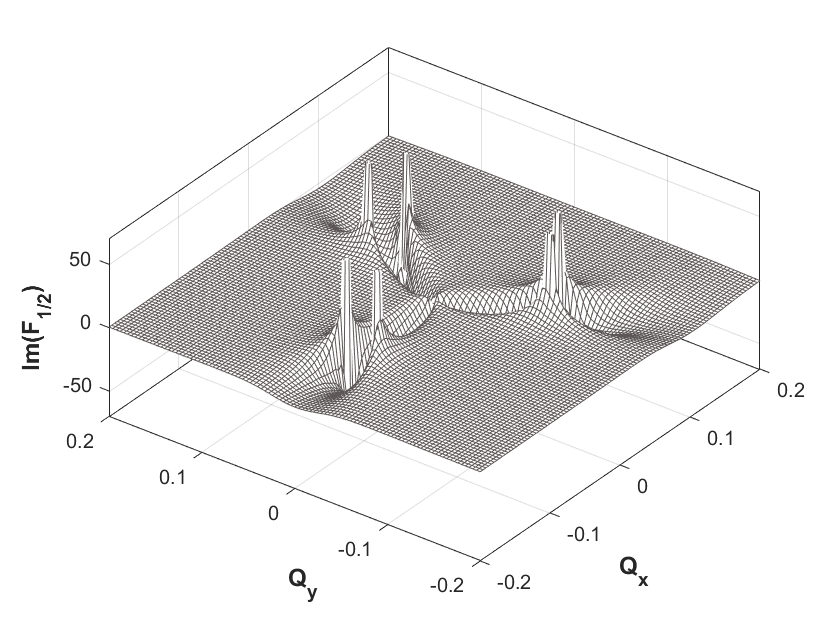}
\centering
\caption{The imaginary part of the first derivative non-adiabatic coupling in $\phi$ direction between the two lower electronic resonant states. The imaginary part diverges at the point of the exceptional point, but not at the central complex conical intersection. The coordinates and the non-adiabatic coupling are presented in dimensionless units.}
\end{figure}

\section{Discussion}
The complex adiabatic potential energy surfaces associated with the electronic resonant states of H$_3$ are well described by the complex PJT model. The enriched topology of the intersecting surfaces includes complex conical intersections, exceptional points and seams of intersections. In the present study, the symmetric mode $Q_s$ is excluded. If however it is included, the points of conical intersections becomes seams of conical intersections with the possibility of more topological features. In example, the seams of bound state conical intersection in Li$_3$ has crossing points, so called heptafurcations, when following the symmetric mode \cite{yark2}. Similar characteristics is possible for the resonant states when following $Q_s$.

At the critical radius $\rho_c$ the exceptional points are found, which also marks a border where the geometric phase is cancelled out. For the complex JT model (see \cite{feu2} and appendix) this critical radius is relatively large, such that the JT expansion probably breaks down. The exceptional points are therefore considered as unphysical for the JT model. The opposite feature is found in the complex PJT model, where the exceptional points are found at a relatively small radius when there is a strong PJT coupling. The physicality of these exceptional points can therefore not be excluded based on the expansion argument. 

In the present study, the three lowest electronic resonant states above the potential of the H$_3^+$ ion potential are considered. There exist however higher excited resonant states, \textit{i.e.} a Rydberg series of resonant states converging to the excited ion limits. These are expected to have similar behaviour to the states studied here and for a complete picture these should be included.

In a theoretical study on ion-pair formation in electron recombination with H$_3^+$ (H$_3+e \to \textrm{H}_2^++\textrm{H}^-$), the wave packet dynamics on the three resonant states of H$_3$ were investigated \cite{roos}. The lowest electronic resonant state (the $A_1$ state in $C_{2v}$ symmetry) of H$_3$ is diabatically correlated with the ion pair limit at infinity. The electronic couplings involved in the triple intersection were estimated from the coefficients in the configuration interaction calculation (\textit{i.e.}  the bound state calculation) and with no regards to the JT or the PJT effects. Possible effects originating from the complex PJT model is however complicated to reveal in the case of a process such as ion-pair formation. The resonant states cross and interact with the series of bound Rydberg states with potentials below the ground state of the ion. This will cause many avoided crossings that will influence the quantum molecular dynamics. Previous wave-packet dynamics study with a simplifed model excluding the Rydberg series shows a pronounced effect of the second order continuum interaction among the resonant states in the regions where the the non-adiabatic coupling is strong \cite{royal}. 
To investigate the potential importance of the PJT effect among the resonant states, we propose to study a process such as resonant vibrational excitation $\textrm{H}_3^+ +e^- \to \textrm{H}_3^{*} \to \textrm{H}_3^{+*}+e^-$. In this process, the electron is temporally captured into the electronic resonant states in the vicinity of the central conical intersection before the system autoionizes leaving the ion in a different ro-vibrational state. The system will here not probe the region where the resonant states cross the potential of the ion and interact with the bound Rydberg states. Additionally, by projecting onto different vibrational states of the ion, the effect of the symmetry distortion can be investigated.

\section{Appendix}
The features present in the topology for the two lowest states of the complex PJT model is also present in the $2 \times 2$ complex JT model. Even though a quantitative fit cannot be obtained, as seen in Fig. 2 and Fig. 3, the JT model captures the topological features and analytic expressions are available. These analytical expressions are given in this appendix. Excluding the $A$ state from the analysis, \textit{i.e.} setting $\alpha=0$, the limit of a pure JT interaction is obtained. The interaction among the degenerate $E$ components is described by the $2\times 2$ potential

\begin{equation}
V_{nm}^d= \big(\varepsilon_E + \frac{1}{2}\omega \rho^2\big) \delta_{nm} + kJ_{nm}^k+gJ_{nm}^g,
\end{equation}
where $J_{nm}^k$ and $J_{nm}^g$ denotes the non-zero elements of the matrices in \eqref{J1} and \eqref{J2}. The complex adiabatic potential energy surfaces are 
\begin{equation}
V_{1/2}= \varepsilon_E + \frac{\omega}{2}\rho^2  \mp \rho\sqrt{k^2 + g^2 \rho^2 + 2 k g \rho \cos(3\phi)}. \label{capes}
\end{equation}
These complex adiabatic potential energy surfaces were studied in detail in \cite{feu2}, but we provide further characteristics of the topology via the non-adiabatic coupling and the geometric phase. By introducing the complex parameter

\begin{equation}
\theta = \textrm{arctan}\bigg(\frac{k\sin (\phi) - g \rho \sin(2\phi) }{k\cos (\phi) + g\rho \cos (2\phi)}\bigg), \label{eqTheta}
\end{equation}
the right eigenvector matrix can be expressed as

\begin{equation}
\textbf{T}(\theta) =   \begin{pmatrix}
  \cos(\theta/2) & \sin(\theta/2)  \\
  \sin(\theta/2) & -\cos(\theta/2) \\
  \end{pmatrix}.
\end{equation}
with left eigenvector matrix $\textbf{T}^T$. Since only the real part of the angle $\theta$ in \eqref{eqTheta} is multivalued, the complex generalisation does not introduce issues concerning the single valuedness of the eigenvector matrices. Single valued biorthogonal states can be constructed by adding a phase, as $\textbf{T}_s=e^{-i Re(\theta)/2}\textbf{T}$ with dual eigenvector matrix defined as $\tilde{\textbf{T}}_s=e^{i Re(\theta)/2}\textbf{T}^T$. Similarly as in the bound state situation this induces a gauge transformation \cite{mead} on the complex non-adiabatic coupling. The (single valued) first derivative non-adiabatic coupling reads

\begin{equation}
\textbf{F}^s = \frac{1}{2}  \begin{pmatrix}
  -i \nabla_{\textbf{Q}} \textrm{Re}(\theta) & \nabla_{\textbf{Q}} \theta  \\
  -\nabla_{\textbf{Q}} \theta & -i \nabla_{\textbf{Q}} \textrm{Re}(\theta)\\
  \end{pmatrix}. \label{eqF} \\
\end{equation}
The geometric phase can be evaluated by integrating the diagonal elements of the non-adiabatic couplings around a close loop around the conical intersection \cite{zwan, jorg},

\begin{equation}
\tau = \frac{1}{2} \mathrm{Re} \Big( \oint_C \nabla_{\textbf{Q}} \theta \ d\textbf{Q} \Big).
\end{equation}
The geometric phase is obviously real. The gradient of the complex adiabatic-diabatic transformation angle $\theta$ in $\phi$ and $\rho$ direction evaluates like in the case for bound states \cite{jorg}

\begin{align}
 \nabla_{\textbf{Q}} \theta &= \frac{-\frac{g}{k} \sin(3\phi)}{1 + (\frac{g}{k})^2 \rho^2 + 2\frac{g}{k}\rho \cos(3\phi)} \widehat{\rho} \nonumber \\
&+ \frac{1}{\rho} \frac{1 - 2(\frac{g}{k})^2\rho^2 - \frac{g}{k}\rho \cos (3\phi)}{1+(\frac{g}{k})^2 \rho^2 + 2\frac{g}{k}\rho \cos(3\phi)} \widehat{\phi}
\end{align}
 but with $k$ and $g$ being complex valued. 

To unfold the topology and the  points of intersections it is convenient to rewrite the the JT potential as

\begin{align}
\textbf{V} &= \big(\varepsilon_E + \frac{1}{2}\omega \rho^2 \big)\textbf{I} \label{Vu} \\
&+ k \rho \big(\cos (\phi)+\frac{g}{k}\rho \cos(2\phi)\big) \begin{pmatrix}
1 & \lambda \\
\lambda & -1
\end{pmatrix} \nonumber
\end{align}
in terms of the complex parameter
\begin{equation}
\lambda = \frac{k\sin (\phi) - g \rho \sin(2\phi) }{k\cos (\phi) + g\rho \cos (2\phi)}.
\end{equation}
Diagonalising $\textbf{V}$ gives $V_{1/2}= \varepsilon_E + \frac{1}{2}\omega \rho^2 \mp u$, with
\begin{equation}
u = \rho k \big( \cos(\phi)+\frac{g}{k}\rho \cos(2\phi)\big) \sqrt{1+\lambda^2 } \label{u}.
\end{equation}
Intersections are to be found when $u=0$, and can be analysed for three scenarios.Obviously $u=0$ at $\rho=0$ and the two complex potential energy surfaces intersects for both the real and imaginary parts. In the very near vicinity of $\rho=0$, the linear JT model ($g=0$) applies. In this region $\theta=\phi$ and the geometric phase evaluates to $\tau=\pi$, corresponding to a conventional sign change as in the bound state situation. Reinstate $g \neq 0$, and expanding the non-adiabatic coupling around $\rho=0$, the angular part reads

\begin{equation}
\nabla_{\phi} \theta = \frac{1}{\rho}-3\frac{g}{k}\cos(3\phi)+\mathcal{O}(\rho).
\end{equation}
The diverging term $1/\rho$ is only carried in the real part, and the imaginary part of the non-adiabatic coupling is not singular at $\rho=0$. 

Outer intersection where $u=0$ can be found at radius $\rho_c=|k|/|g|$, where two cases can be distinguished. For real valued $k/g$, three point of intersections are possible when $\cos(\phi)+g/k \rho \cos(2\phi)=0$. The bound state situation with $\textrm{Im}(k)=\textrm{Im}(g)=0$ is a special case and the associated geometric phase for these intersections are $\tau=\pi$, analogous to the bound state JT case \cite{jorg}.

The resonance situation allows for the possibility that $g/k$ is complex manifesting non-hermitian effects. The matrix in \eqref{Vu} is a well known non-hermitian matrix exhibiting exceptional point (or non-hermitian degeneracies) at $\lambda=\pm i$ \cite{heiss}. At an exceptional point, not only the complex eigenvalues intersects, but also the eigenstates coalesce into one, which implies that the eigenvector matrix $\textbf{T}$ is singular and both the real and imaginary parts of the non-adiabatic coupling diverges. The geometric phase evaluates to $\tau=\pi/2$. The exceptional points are further connected by seams, where $\textrm{Re}(V_{1})=\textrm{Re}(V_{2})$ or $\textrm{Im}(V_{1})=\textrm{Im}(V_{2})$ intersects. These are situated at angles given by \cite{feu2}

\begin{equation}
\cos 3\phi = -\frac{1}{\rho}\frac{\textrm{Re}(k)\textrm{Im}(k)+\textrm{Re}(g)\textrm{Im}(g)\rho^2}{\textrm{Re}(k)\textrm{Im}(g)-\textrm{Re}(g)\textrm{Im}(k)}
\end{equation}
which meets at $\rho_c=|k|/|g|$ in the six exceptional points corresponding to $\lambda=\pm i$. Encircling the central intersection for a fixed radius the geometric phase evaluates like the bound state case \cite{jorg} with $\tau=\pi$ for $\rho<|k|/|g|$ while $\tau=2\pi$ for $\rho>|k|/|g|$. The topological features found in the complex PJT model between the two lower electronic resonant states are analogous the the features presented in this appendix for the complex JT model.

\end{document}